\documentclass[a4paper, amsfonts, amssymb, amsmath, reprint, showkeys, nofootinbib, twoside, pre]{revtex4-2}
\usepackage[english]{babel}
\usepackage[utf8]{inputenc}
\usepackage[colorinlistoftodos, color=green!40, prependcaption]{todonotes}
\usepackage{amsthm}
\usepackage{mathtools}
\usepackage{physics}
\usepackage{xcolor}
\usepackage{graphicx}
\usepackage[left=23mm,right=13mm,top=35mm,columnsep=15pt]{geometry} 
\usepackage{adjustbox}
\usepackage{placeins}
\usepackage[T1]{fontenc}
\usepackage{lipsum}
\usepackage{csquotes}
\usepackage[pdftex, pdftitle={Article}, pdfauthor={Author}]{hyperref}
\usepackage{verbatim}
\begin{document}
\title{Direct Laser Acceleration of Bethe-Heitler positrons in laser-channel interactions}

\author{Bertrand Martinez}
    \email[Correspondence email address: ]{bertrand.martinez@tecnico.ulisboa.pt}
    \affiliation{Golp/Instituto de Plasma e Fus\~{a}o Nuclear, Instituto Superior T\'{e}cnico, Universidade de Lisboa, 1049-001 Lisbon, Portugal}

\author{Robert Babjak}
    \affiliation{Golp/Instituto de Plasma e Fus\~{a}o Nuclear, Instituto Superior T\'{e}cnico, Universidade de Lisboa, 1049-001 Lisbon, Portugal}
    \affiliation{Institute of Plasma Physics, Czech Academy of Sciences, Za Slovankou 1782/3, 182 00 Praha 8, Czechia}

\author{Marija Vranic}
    \affiliation{Golp/Instituto de Plasma e Fus\~{a}o Nuclear, Instituto Superior T\'{e}cnico, Universidade de Lisboa, 1049-001 Lisbon, Portugal}

\begin{abstract}
Positron creation and acceleration is one of the major challenges for constructing future lepton colliders. 
On the one hand, conventional technology can provide a solution, but at a prohibitive cost and scale.
On the other hand, alternative, reduced-scale ideas for positron beam generation could bring this dream closer to reality. 
Here we propose a novel plasma-based positron acceleration method using a powerful laser propagating through a dense and narrow plasma channel.
A large amount of electrons is injected within the channel during laser propagation.
This electron loading creates static fields in the plasma, enabling positrons to be guided transversely while they directly gain energy from the laser field itself.
Within this context, we present a  theoretical model to describe how the laser injects the electrons and estimate the beam-loaded effective electron density. 
We validate our theoretical predictions through Quasi-3D PIC simulations and demonstrate the robustness of this guiding and direct laser acceleration process for positrons.
Our approach could pave the way for testing  this new positron acceleration scheme at ELI-Beamlines, showcasing unprecedentedly high average energy gain rate of a few TeV/m.
The fireball jet produced contains GeV-level electrons, positrons, and x-rays, opening the path towards potential laboratory astrophysics experiments using these beams.
\end{abstract}

\maketitle

\section*{Introduction}

Investigation into positron acceleration stands as a pivotal endeavor of fundamental physics with profound implications not only for future colliders, but also for laboratory astrophysics.
In particular, creating jets with GeV-level electrons, positrons and gamma-rays will mark a major milestone toward exploring the propagation of fireball beams in a plasma medium, and thus the physics of Gamma-Ray Bursts.  

High-energy positrons can be produced in state-of-the-art radio-frequency accelerators.
A promising concept to further boost the peak energy of existing positron beams was introduced experimentally in the 2000s~\cite{PRLNg2001,PRLHogan2003,PRLBlue2003,PRLMuggli2008}, and later complemented~\cite{PRSTABKimura2011,NatCorde2015,NCGessner2016,SRDoche2017,PRLLindstrom2018}.
This approach leverages a plasma stage, capable to withstand accelerating fields of $10 \, \rm GV/m$, a thousand times larger than conventional accelerators ($10 \, \rm MV/m$).
These accelerating fields can be formed in the blowout regime of plasma wakefield acceleration.
The main challenge is to guide the positrons in the direction transverse to their propagation while maintaining a low emittance.
One solution is to tailor the conditions of the interaction to induce the formation of a dense electron filament that focuses  positrons transversely.
For instance, this can be achieved using a Laguerre-Gauss laser~\cite{PRLVieira2014,PoPYu2014}, a hollow electron beam driver~\cite{PRLJain2015}, a non-neutral electron-positron driver~\cite{PRABSilva2023}, an externally injected positron beam~\cite{PRALiu2023,arXivZhou2022}, or with a plasma column~\cite{PRABDiederichs2019}.
While this research is conducted in particle accelerators, it became recently a topic of investigation in high-power laser facilities~\cite{RLECheriaux2018,HPLSEBromage2019,CLEOPapadopoulos2019,OEYoon2019}.
With upcoming high-power lasers (4-5 PW), positrons created by the linear Breit-Wheeler process~\cite{PRBreit1934} are expected to be accelerated by direct laser acceleration~\cite{CPHe2021} or by self-generated longitudinal plasma fields~\cite{PRLSugimoto2023}.
With lasers above 10 PW, it is proposed to create nonlinear Breit-Wheeler positrons and accelerate them in vacuum~\cite{SRVranic2018}, via direct laser acceleration~\cite{PRABMartinez2023,PoPMaslarova2023}, or in a recovering plasma wakefield~\cite{PRRLiu2022}.

The main constraint in the realm of positron acceleration research is the limited availability of large accelerator facilities globally.
Although this issue can be mitigated by utilizing high-power lasers, which are not only more compact but also more affordable, there is a notable scarcity of studies focusing on this approach in the existing literature.
The lack of research in this area underscores the need for further investigation to advance our understanding and capabilities in the domain of positron acceleration.
The development of new theoretical models and their validation through accurate simulation data is crucial to bring forward potential experiments at high-power laser facilities, finding applications in the domain of laboratory astrophysics.


\begin{figure*}
    \centering
    \includegraphics[width=0.80\textwidth]{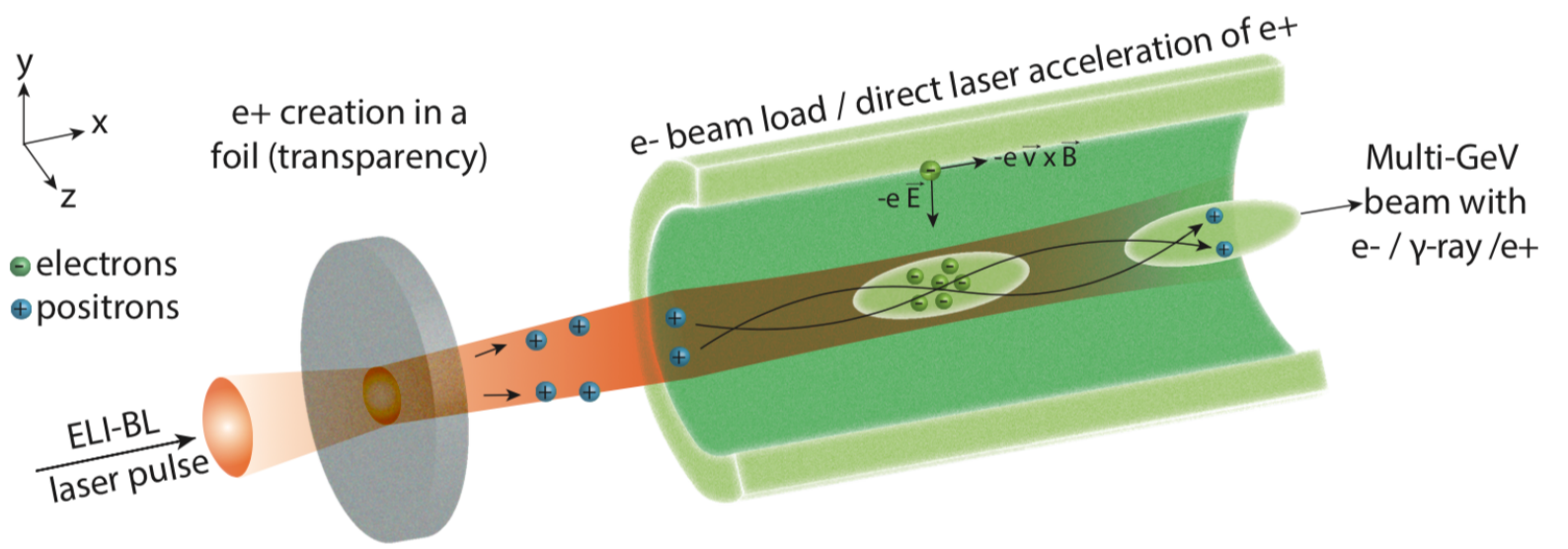}
    \caption{Setup for Direct Laser Acceleration (DLA) of Bethe-Heitler positrons. The $10\, \rm PW$ ELI-Beamlines laser first interacts with a thin Aluminum foil in the Relativistically Self-Induced Transparency regime and creates positrons. In the second step, it co-propagates, along with some of the positrons, in a preformed plasma channel. A dense electron filament is progressively built up at the center of the channel via the interplay of the two laser field components. This enables to guide positrons over a millimeter distance while they gain energy directly in the strong laser field, with an average accelerating energy gain rate on the order of a few TeV/m.
    }
    \label{fig:setup}
\end{figure*}

In this work, we investigate Direct Laser Acceleration (DLA) of Bethe-Heitler positrons in laser-channel interactions.
Since DLA of electrons was introduced~\cite{PoPPukhov1999,PRLGahn1999}, it has been studied theoretically in various conditions~\cite{PRLArevief2012,PoPKhudik2016, PREGong2020, NJPJirka2020, PREWang2021, PRABLi2021, NJPYeh2021,PRLBabjak2024,PREValenta2024}.
In addition, recent experiments highlight that it holds a potential to generate relativistic ($100' \rm s$ of MeV) and high charge ($100' \rm s$ of nC) electron beams~\cite{NJPHussein2021,SRShaw2021}, that are already employed to drive bright sources of secondary particles as X-rays and neutrons~\cite{MRERosmej2021,NCGunther2022}.
With DLA, electrons gain energy in the laser field itself which amplitude is on the order of $\simeq 10 \, \rm TV/m$, a thousand times larger than in a typical plasma wakefield.
For high-power lasers with a peak intensity of $10^{21} \, \rm W cm^{-2}$ and an adequate focal spot, electrons are expected to reach peak energies of 10 GeV~\cite{PRLBabjak2024}.
In the transverse direction, they are confined by the self-generated plasma fields.
With their opposite charge, positrons are transversely expelled by the same fields and therefore cannot be accelerated.
Recent but yet unexplored developments suggest that positron guiding can be achieved in DLA either using two counter-propagating lasers in a channel~\cite{CPHe2021}, or by forming a dense electron filament on the laser axis~\cite{PRABMartinez2023}.

In the setup we introduce in Fig.~\ref{fig:setup}, a high-power laser (as will be soon available at ELI-Beamlines $\sim 10 \, \rm PW $)  interacts with a thin aluminum foil ($240 \, \rm nm $).
The pulse intensity is high enough ($\sim 10^{22} \, \rm W cm^{-2}$) to induce gamma-ray production via nonlinear Inverse Compton scattering~\cite{JMPReiss1962} and positron creation through the Bethe-Heitler process~\cite{RMPMotz1969} in the plasma.
Due to its ultra-fast heating and expansion, the target becomes transparent to the laser in the so-called Relativistically-Self-Induced Transparency regime~\cite{PFKaw1970}.
After this interaction, the laser and some of the positrons travel together in a preformed plasma channel of near-critical density placed on the laser propagation axis.
During the first part of laser propagation in the channel, the self-generated plasma fields are established to be defocusing for positrons and focusing for electrons~\cite{PoPKostyukov2004,PoPGordienko2005}, so a significant fraction is lost.
In the meantime, electrons are injected from the walls of the channel to its center by the interplay of the transverse electric and transverse magnetic components of the laser field, as sketched in Fig.~\ref{fig:setup}.
The amount of electrons loaded increases as the laser propagates further in the channel, progressively damping the transverse force expelling positrons.
When the charge density of electrons loaded becomes larger than the background ion charge density, the transverse force acting on positrons changes sign and they can be guided. 
They perform low-frequency betatron oscillations in the loaded fields of the plasma channel, but are also subject to high-frequency oscillations in the laser field.
The oscillatory motion of a given positron can be resonantly amplified when the two frequencies match in its co-moving frame.
From this resonance, there is a large energy transfer from the laser to the positron and it gains momentum in the laser propagation direction.
This acceleration process is the generalization of Direct Laser Acceleration for positrons.

In the first section of this article, we derive estimates describing the process of electron injection and how it scales with the initial plasma and laser conditions.
Leveraging this result, we present an energy scaling law for Direct Laser Acceleration of positrons, including a resonance condition.
The second section is dedicated to Quasi-3D PIC simulations of positron acceleration.
We validate the trend predicted by our estimate for electron beam loading, as well as our assumption of injection from the walls of the channel.
The acceleration process of positrons is analysed in detail using individual trajectories and a characteristic feature of DLA known as a forking structure is observed in their simulated spectrum~\cite{PPCFShaw2018,SRShaw2021,PRABKing2021}.
Finally, we discuss a possible path to increase the conversion efficiency from laser energy to positrons and how DLA of positrons compares to other approaches.
We conclude with an assessment of experimental feasibility and potential applications.

\section{Theoretical model of positron acceleration \label{sec:theory}}

In this section, we introduce a theoretical model to describe Direct Laser Acceleration of positrons in a plasma channel.
Our first and main result is to provide a quantitative estimate for the charge density of the electrons loaded by the intense laser traveling in the plasma channel.
From this general result derived as a function of the laser and plasma conditions, we bring forward a model describing the process of Direct Laser Acceleration for positrons.

\subsection{Injection of electrons and beam loading}

We consider a laser propagating in direction $x$ and linearly polarised in direction $y$.
It has a frequency $\omega_L$,  a wave number $k_L$, a peak field amplitude $E_L$ and we denote $n_c$ the critical density associated.
It travels in a preformed plasma channel characterized by a radius denoted $r_c$.
We use the notation $n_p$ for the density in the center of the channel and $n_w$ for the density in the walls of the channel.
In other words, the transverse density profile of the plasma channel is given by $n(r) = n_p + (n_w - n_p ) (r/r_c)^\alpha$, where $r$ is the radial coordinate defined as $r^2=y^2+z^2$, and $\alpha$ is an integer describing the steepness of the profile ($\alpha=2$ for a parabolic profile).
The laser duration $\tau_L$ is larger than the plasma period.

As the laser travels in the plasma channel, the ponderomotive force expels electrons and a blowout cavity is formed.
Behind this cavity, electrons are pushed transversely toward the center by the electric field component of the laser $-e\mathbf{E}$, and rotated in the laser propagation direction by its magnetic field component $-e\mathbf{v}\times \mathbf{B}$, where $e$ is the electron charge and $v$ the electron velocity.
To evaluate the amount of electrons loaded, we have to simplify the problem and formulate an additional hypothesis.
We justify this assumption a posteriori in our analysis of PIC simulations performed in section~\ref{sec:simulations}.
Although the laser focusing and defocusing dynamic is complex, we assume that its typical transverse extent can be approximated by the blowout radius of the interaction, denoted $r_b$.
To derive this radius, we balance the ponderomotive force of the laser as defined in this work~\cite{PREQuesnel1998}  with the electrostatic force created in the channel, provided by Poisson equation $ \nabla E = \rho / \epsilon_0$ where $\rho(r) = -e[n_p + (n_w - n_p ) (r/r_c)^\alpha]$ is exactly the generic plasma channel profile we model.
We define the dimensionless laser field amplitude $a_L=eE_L/m\omega_Lc$, where $m$ is the electron mass and $c$ is the light velocity.
In the calculation of the ponderomotive force, we assume that the average Lorentz factor of electrons is $\langle \gamma \rangle \simeq a_L$. The blowout radius is thus a solution of
\begin{equation} \label{eq:radiusinjection}
     k_p^2 r^2 \left[1+\frac{1}{\alpha+1}\left(\frac{n_w}{n_p} - 1\right)\left( \frac{r}{r_c}\right)^\alpha \right] = \frac{a_L}{2}
\end{equation}
where $k_p$ is the plasma wave number associated with the central channel density $n_p$.
This equation is a generalisation of the established expression of the blowout radius from a constant density plasma ($n_w=n_p$), to a plasma with an increasing profile ($n_w > n_p$).
It accounts for arbitrary variations of the density profile and therefore has a broad range of validity.
For instance, it encompasses the case of plasma channels with a parabolic profile ($\alpha=2$) and also the case of a plasma tube with an infinitely steep profile  ($\alpha\rightarrow \infty$ where $r_b \rightarrow r_c$).
While analytical solutions can be derived for specific values of the exponent $\alpha\in(0,1,2,4,6)$, it is worth noting that the expressions become cumbersome to handle analytically.

\begin{figure}
    \centering
    \includegraphics[width=0.48\textwidth]{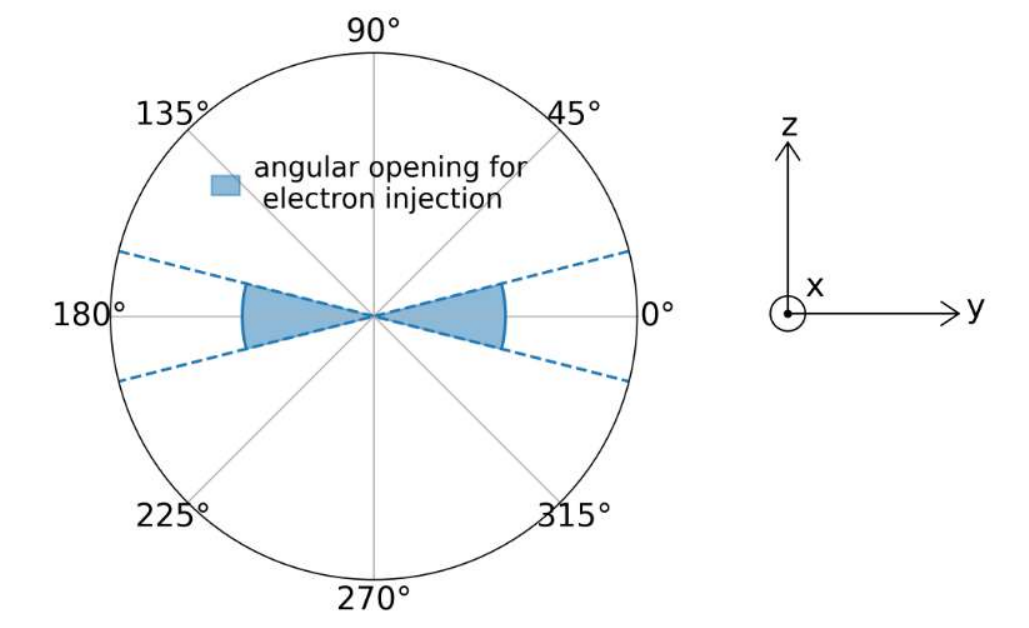}
    \caption{Electrons are injected mostly along direction $y$, as the laser is linearly polarised in this direction and propagates along the $x$ axis.
    In the model, we assume electrons are injected from an angle with an opening of $\sim 1 \, \rm rad$.
    }
    \label{fig:angle_inj}
\end{figure}

At the blowout radius $r_b$, the plasma density is denoted $n_b = n(r_b)$.
We assume that at every laser cycle, all electrons within the local skin depth are injected in the channel.
This means that all electrons between the two cylinders of radius $r$ and $r+\delta$ are injected, where the skin depth $\delta$ is defined as $\delta=(c/\omega_L)(n_c/n_b)^{1/2}$.
For a laser linearly polarised in direction $y$, the electric field component of the laser is actually not always perpendicular to the surface, so that electrons are not injected from all radial angles in the channel (namely on a $2\pi$ angle), but rather on a smaller opening angle in the $\pm y$ direction, that we approximate to $\simeq 1 \, \rm rad$, as illustrated in Fig.~\ref{fig:angle_inj}
With this reasoning, we can obtain a rate of electrons injected per laser period $dN/dx \simeq r_bn_b\delta$, assuming $\delta \ll r_b$.
However, we are interested in estimating the density of electrons loaded in the plasma channel, denoted $n_e$.
To derive this, we assume that at any time $t$, the electrons loaded fill on average a cylinder of volume $\pi r_b^2 v_g t$, where $v_g$ is the group velocity of the laser and the radius is chosen to be the blowout radius.
We deduce that both the number of electrons and the volume they fill increase linearly with time.
When we derive the ratio of these two quantities, namely the density of electrons loaded $n_e$, the linear time dependence simplifies and we deduce that the average density of electrons loaded per laser cycle is constant
\begin{equation} \label{eq:density_loaded}
    n_e = \frac{1}{\pi r_bk_L} (n_b n_c)^{1/2}
\end{equation}
For a laser with $N$ cycles, this average density is increased by a factor $N$.
We stress that this expression is derived for any plasma channel with a transverse density profile $n(r) = n_p + (n_w - n_p ) (r/r_c)^\alpha$, where $r_c$ is the channel's radius, $n_p$ is the central channel density, $n_w$ is the wall density and $\alpha$ the steepness of the profile.
In Eq.~\eqref{eq:density_loaded}, the blowout radius $r_b$ is given for this exact profile by Eq~\eqref{eq:radiusinjection}, and the density $n_b$ is the channel's density at this radius, namely $n_b = n(r_b)$ for this same profile.
We underline that the two quantities $r_b$ and $n_b=n(r_b)$ depend on the steepness of the plasma profile $\alpha$ as $r_b$ is a solution of Eq.~\eqref{eq:radiusinjection}.
Therefore, our estimate in Eq;~\eqref{eq:density_loaded} has a broad validity range since it provides a scaling of electron beam loading with respect to the full plasma channel profile and the laser intensity.

\subsection{Model for Direct Laser Acceleration of positrons}

\begin{figure}
    \centering
    \includegraphics[width=0.48\textwidth]{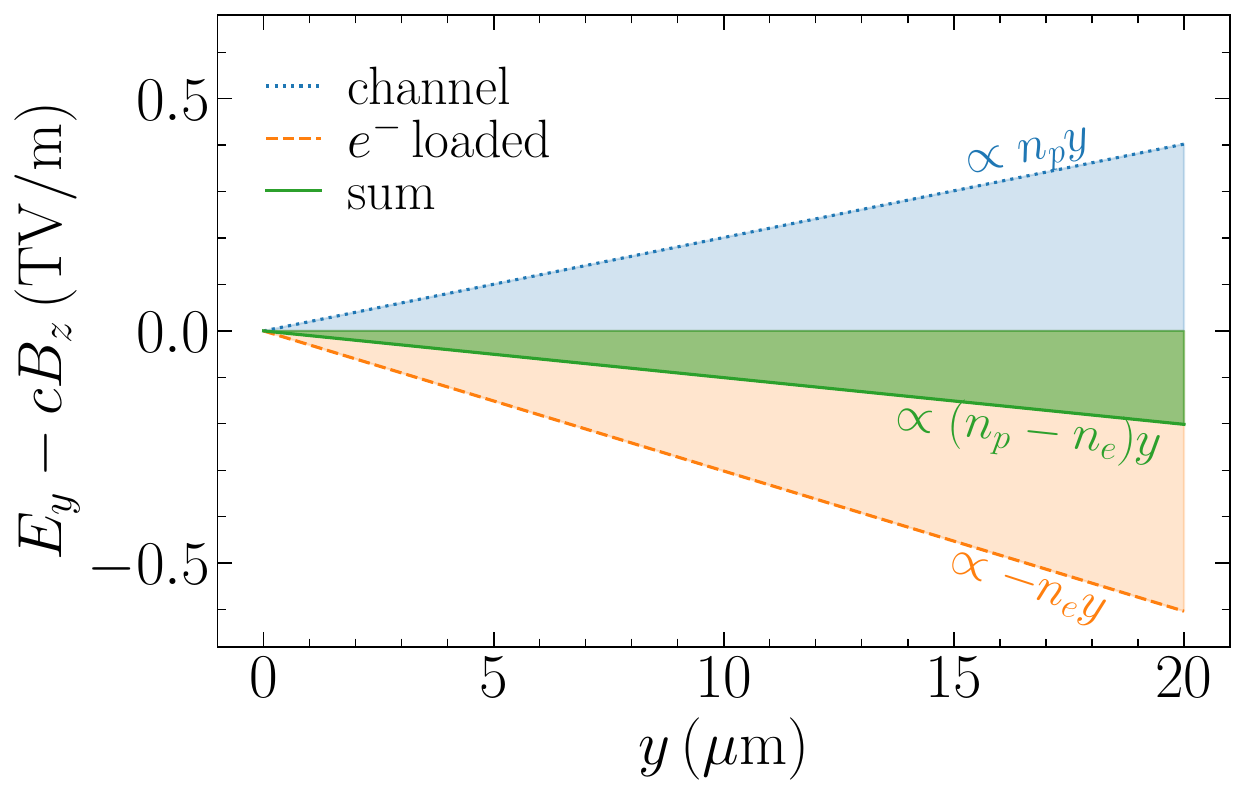}
    \caption{Transverse fields in a plasma channel as a function of the transverse distance to the axis $y$. When the resulting field is positive, electrons are focused and positrons defocused. On the contrary, if the resulting field is negative, electrons are expelled and positrons are guided. We display two contributions: the plasma ion cavity with an ion density $ n_p=10^{19} \, \rm cm^{-3}$ (dotted blue line) and the electron beam loading with density $n_e=1.5\times 10^{19} \, \rm cm^{-3}$ (dashed orange line). The resulting fields (green line) are negative as the net charge density is negative.
    }
    \label{fig:sketch_theory} 
\end{figure}

In the established framework of Direct Laser Acceleration, the laser expels electrons via its ponderomotive force, forming an ion cavity with a positive radial electric field $E_y$ and an azimuthal magnetic field $B_z$, such that  $E_y -cB_z \propto n_p y$.
This positive component is illustrated schematically in Fig.~\ref{fig:sketch_theory} with the dotted blue line.
The electrons loaded at the center of the channel also create static fields.
The contribution of relativistic electrons to the transverse static fields is smaller by a factor $1/\gamma^2$ as compared to the non-relativistic ones.
We therefore assume the transverse fields are the same as for a non-relativistic electron beam of density $n_e$, namely $E_y-cB_z\propto -n_e y$, .
This new and negative contribution is represented by the dashed orange line in Fig.~\ref{fig:sketch_theory}.
While we expect these electrons to be left behind the laser after some distance, new electrons are continuously injected, keeping the overall loaded electron density constant in the channel.
The contribution of the relativistic electrons loaded is expected to vanish as it scales with $n_e/\gamma^2$ where $\gamma$ is the Lorentz factor of the fast electrons.
Accounting for the two contributions of the ion cavity and the non-relativistic electron beam loaded, the resulting transverse field is
\begin{equation} \label{eq:Fperp}
    E_y - c B_z = \frac{e}{\epsilon_0} \frac{y}{2} (n_p - n_e)
\end{equation}
Depending on the amount of electron beam loading, the resulting transverse field can be either positive or negative.
With a negligible electron beam loading ($n_e \ll n_p$), the transverse force on positrons is positive and they are expelled radially away from the channel axis, as established by previous works~\cite{PoPKostyukov2004,PoPGordienko2005}. 
However, when the electron beam loading density overcomes the background ion density ($n_e > n_p$), the transverse force on positrons reverts its sign, becomes negative and thus acts toward a positron guiding.
The latter case is illustrated in Fig.~\ref{fig:sketch_theory} with the green line.
It is worth highlighting that our expression for the resulting transverse field, is obtained with intuitive arguments.
However, the same expression has already been formally derived in the context of beam loading for plasma wakefield acceleration~\cite{PRLVieira2014,arXivZhou2022}.

The transverse force experienced by an electron or a positron of charge $q=\pm e$ in a plasma channel with beam loading is deduced from Eq.~\eqref{eq:Fperp}.
We denoted it $F_c$ and it can be expressed as a function of an effective plasma frequency $\omega_{\mathrm{eff}}$
\begin{align}
    F_{c} & = \frac{q}{e}  \frac{y}{2} m\omega_{\mathrm{eff}}^2 \nonumber \\
    \text{with }& \omega_{\mathrm{eff}}^2 = \omega_p^2(1-n_e/n_p) \label{eq:omegape}
\end{align}
In the limit of low electron beam loading ($n_e \ll n_p$), the effective plasma frequency is exactly the plasma frequency and we recover the established transverse force of an ion cavity  for an electron $F_{c}=-m\omega_p^2y/2$~\cite{PoPKostyukov2004,PoPGordienko2005}.

The laser is modeled as a plane wave propagating in direction $x$, linearly polarised in direction $y$ as $E_y^{(L)}=E_L \cos\omega_L(t-x/v_\phi)$.
It has a magnetic field component $B_z^{(L)}=E_y^{(L)}/v_\phi$, where $v_\phi>c$ is the phase velocity of the plane wave in the plasma of density $n_p$.
The equation of motion and energy conservation can be written as
\begin{align}
\frac{d p_x}{dt} & = q v_y B_z^{(L)} \label{eq:dpxdt} \\
\frac{d p_y}{dt} & =  q \left( E_y^{(L)} - v_x B_z^{(L)} \right)  + \frac{q}{e}  \frac{y}{2} m\omega_{\mathrm{eff}}^2 \label{eq:dpydt} \\
mc^2\frac{d \gamma }{dt} & = q v_y \left( E_y^{(L)} + \frac{1}{e}  \frac{y}{2} m\omega_{\mathrm{eff}}^2 \right) \label{eq:dgdt}
\end{align}
We note that the transverse force induced by the channel fields includes the effect of beam loading via the effective plasma frequency $\omega_{\mathrm{eff}}$, as defined in Eq.~\eqref{eq:omegape}.
With Eq.~\eqref{eq:dpxdt} and Eq.~\eqref{eq:dgdt} we find that there is an invariant of motion ($dI_0/dt = 0$) with $I_0$ defined as
\begin{equation} \label{eq:Invariant}
    I_0 = \gamma - \frac{v_\phi}{c}\frac{p_x}{mc} -\frac{q}{e} \frac{\omega_{\mathrm{eff}}^2y^2}{4c^2}
\end{equation}
From Eq.~\eqref{eq:dpydt} and Eq.~\eqref{eq:dgdt}, we can deduce the equation of motion for the transverse coordinate of the particle $y$.
This derivation is a generalisation of Ref~\cite{PoPPukhov1999} and leads to
\begin{align}
\frac{d^2y}{dt^2} + \omega_\beta^2 y = -q & \left[ \left(\frac{1}{c}\frac{dy}{dt}\right)^2 - \left(1-\frac{v_x}{v_\phi}\right) \right]\frac{E_y^{(L)}}{m\gamma} \nonumber \\
& -\frac{q}{e} \left(\frac{1}{c}\frac{dy}{dt}\right)^2 \frac{\omega_{\mathrm{eff}}^2}{2\gamma}y \label{eq:harmonic}
\end{align}
where the betatron frequency $\omega_\beta$ is defined as
\begin{equation}
\omega_\beta = \left\{
    \begin{array}{ll}
        \frac{\omega_p}{\sqrt{2\gamma}} \sqrt{1-\frac{n_e}{n_p}} \mbox{ for } e^- \mbox{ if } n_e<n_p \\
        \frac{\omega_p}{\sqrt{2\gamma}} \sqrt{\frac{n_e}{n_p}-1} \mbox{ for } e^+ \mbox{ if } n_e>n_p
    \end{array}
\right. \label{eq:omegab}
\end{equation}
This definition evidences in what conditions electrons and positrons are allowed to oscillate resonantly.
Electrons can experience a resonant acceleration only if the beam loading remains smaller than the initial plasma density ($n_e<n_p$).
On the contrary, positrons can experience a resonant acceleration only if the beam loading becomes larger than the initial plasma density ($n_e>n_p$).

The method to derive Eqs.~\eqref{eq:dpxdt}-\eqref{eq:harmonic} is strictly similar to the one already established for DLA of electrons.
Indeed, for the case of an electron ($q=-e$) and without beam loading ($n_e=0$), we recover the established definition for the invariant of motion $I_0$ as defined in Ref~\cite{PoPPukhov1999}, the betatron frequency $\omega_\beta^2=\omega_p^2/2\gamma$ and the equation describing the transverse oscillations of the particle~\cite{PoPPukhov1999}.
Formally, the only change we did is to account for beam loading by substituting the plasma frequency $\omega_p$ by an effective plasma frequency $\omega_{\mathrm{eff}}$, as defined in Eq.~\eqref{eq:omegape}.
We can draw an important conclusion from this mathematical similarity.
The scaling laws for DLA of electrons and positrons with beam loading can be generalized from the already established ones by a simple substitution $\omega_p \rightarrow \omega_{\mathrm{eff}}$, provided the resonance condition in Eq.~\eqref{eq:omegab} is fulfilled.
For positrons, namely in the case of high beam loading where $n_e>n_p$, the maximum energy at resonance averaged on a betatron period is~\cite{PoPPukhov1999,PRLArevief2012,PoPKhudik2016}
\begin{equation} \label{eq:scaling_gamma}
\langle \gamma \rangle = \frac{3}{4} I^2 \left(\frac{\omega_L}{\omega_{\mathrm{eff}}}\right)^2 
\end{equation}
In the context of the ELI-Beamlines laser pulse ($150 \, \rm fs$ laser with a $\sin^2$ time envelope) where our setup could be realized, we have estimated the average energy gain rate~\cite{NJPJirka2020}
\begin{equation} \label{eq:scaling_gradient}
\frac{d\gamma}{dx} = \frac{5}{2} \frac{a_L}{\sqrt{I}} \frac{\omega_{\mathrm{eff}}}{\omega_L}
\end{equation}
For high laser intensities, the relativistic positrons emit radiation, lose energy and this should be accounted for in the theory.
This was already calculated in the case of electrons and without beam loading~\cite{NJPJirka2020}.
We verified that it can also be generalised by the substitution of the plasma frequency $\omega_p$ to the effective plasma frequency $\omega_{\mathrm{eff}}$ accounting for beam loading.
The main effect of radiation reaction is that the parameter $I_0$ defined in Eq.~\eqref{eq:Invariant} is not invariant anymore.
The exact evolution of $I$ cannot be derived.
Its steady-state value, averaged over several betatron periods, can however be estimated assuming the particle reached resonance~\cite{NJPJirka2020}
\begin{align} \label{eq:I_rr}
I/I_0 &= \left\{
    \begin{array}{ll}
        & \left[1+3.2\times10^{-8} \frac{\omega_L^2}{\omega_{\mathrm{eff}}^2} \frac{I_0^4}{\lambda_L[\mathrm{\mu m}]}\right]^{-1/4} \\
        & \left[1+2.3\times10^{-8} a_L \frac{I_0}{\lambda_L[\mathrm{\mu m}]}\right]^{-1}
    \end{array}
\right.
\end{align}
where the first expression is valid when the channel fields prevail in the radiation reaction force, and the second as the laser field dominates in the radiation reaction force.

To conclude, we introduced an analytical model to describe Direct Laser Acceleration of positrons in a plasma channel accounting for electron beam loading.
The main outcome is that we obtain an estimate for the electron beam loading as a function of the initial laser and plasma channel conditions.
This is a key theoretical milestone, as it enables to better grasp the conditions of positron acceleration and suggest for the first time a scaling law for the peak energy positrons may reach as they experience DLA.

\section{Quasi-3D simulations of positron acceleration \label{sec:simulations}}

In this section, we detail the results of PIC simulations of Direct Laser Acceleration of positrons.
We first describe the process of beam loading and verify the scaling predicted by our theoretical estimate for a wide range of laser and plasma conditions.
We also discuss the validity of the assumptions of our theoretical model in light of the simulation data.
We finally bring unambiguous proof that positrons can experience Direct Laser Acceleration in Quasi-3D PIC simulations.

\begin{figure*}
    \centering
    \includegraphics[width=0.90\textwidth]{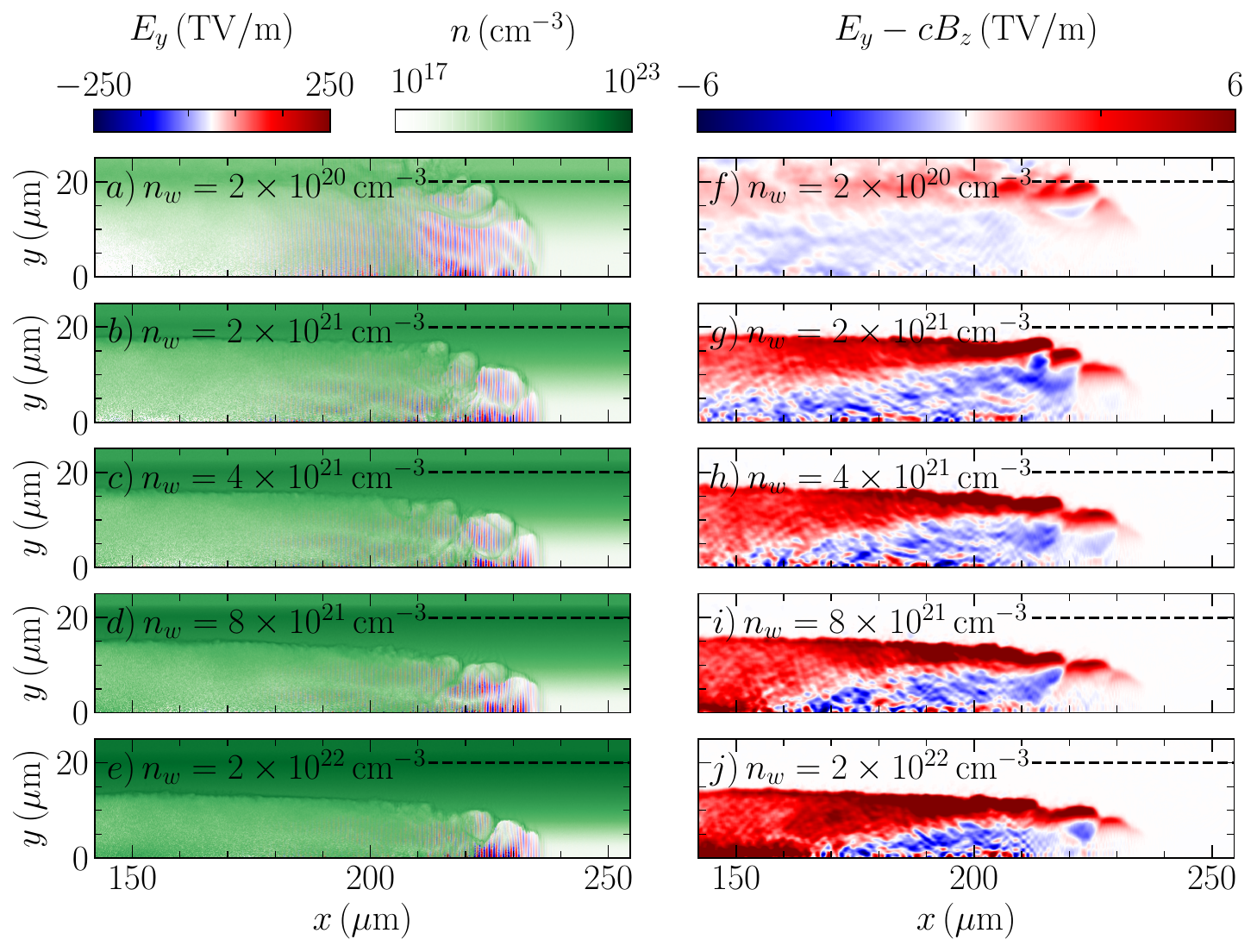}
    \caption{Electron beam loading in laser-channel interactions as a function of the wall density of the channel. Panels a)-e) illustrate the interaction of the laser pulse (red/blue) with the electrons of the channel (green). In panels f)-j), we note that transverse fields are negative in the channel, so positrons can be guided. When moving from the top row to the bottom row, we increase the wall density of the plasma channel in the range $n_w\in (0.2, \, 2, \, 4, \, 8, \, 20) \times 10^{21} \, \rm cm^{-3}$. The data is shown after the peak of the laser has propagated over a distance $L_x=200 \, \rm \mu m$ in the channel and has already propagated in the transparency regime through the aluminum foil placed at $x=0 \, \rm \mu m$.
    The process is robust with respect to the wall density, so the channel could be formed with a capillary discharge or with another laser.
    }
    \label{fig:beam_loading}
\end{figure*}

\subsection{PIC simulations with 10 PW class lasers \label{sec:sim_params}}

The setup in Fig.~\ref{fig:setup} is modeled with the Particle-In-Cell code Osiris~\cite{IPFonseca2002}.
The laser has a wavelength of $1 \, \rm \mu m$, propagates in the x direction, and is linearly polarised in direction $y$.
The electric field of this pulse has a Gaussian temporal profile, with a duration of $150 \, \rm fs$ (FWHM) and a Gaussian transverse profile of $5.3 \, \rm \mu m$ (FWHM).
Its peak intensity is $5 \times 10^{22} \, \rm Wcm^{-2}$, corresponding to a normalised field amplitude $a_L=200$, a peak power of $9 \, \rm PW$ and an energy of $1.3 \, \rm kJ$.
We initialise the thin foil as a fully ionized Aluminum plasma with a thickness of $240 \, \rm nm$.
We assume it is pre-expanded by the laser pre-pulse and has a charge density of $3\times 10^{22} \, \rm cm^{-3}$.
We also initialise an exponential pre-plasma in front of it, with a scale length of $500 \, \rm nm$.
The preformed plasma channel is placed right after the thin foil.
It has a transverse profile given by $n(r)=n_p+(n_w-n_p)(r/r_c)^\alpha$, where $n_p=10^{18} \, \rm cm^{-3}$ is the central density in the channel, $n_w=4 \times 10^{21} \, \rm cm^{-3}$ is the wall density of the channel, $r_c=20 \, \rm \mu m$ is the channel's radius and $\alpha=5$ is the steepness of the profile.
The density profile at the beginning of the plasma channel is smooth with a $10 \, \rm \mu m$ linear gradient increasing from $0$ to the value given by the transverse profile above.

A correct description of laser guiding in a plasma channel requires 3D simulations, that are out of reach for modern supercomputers for the required propagation distances.
To overcome this issue, we perform simulations with the Quasi-3D simulation mode of the PIC code OSIRIS~\cite{JCPDavidson2015}, that correctly models 3D pulse focusing at a cost of the same order as a 2D algorithm.
In the Quasi-3D approach, the fields are represented in cylindrical coordinates $(x,r,\theta)$ using a Fourier decomposition in angular modes~\cite{JCPDavidson2015}.
The first two modes account for the axisymmetric self-generated channel fields (mode m = 0) and for the non-axisymmetric linearly polarized laser field (m = 1).

The simulation domain has a size of $137.5\times80 \, \rm \mu m^2$, with spatial steps $dx=dr=16 \, \rm nm$ and a time step of $26.6 \, \rm as$.
The laser propagates on a total distance of 1 mm in the plasma channel, which corresponds to $1.28\times 10^5$ time steps.
We introduce $L_x$ as the distance travelled by the peak of intensity in the channel.
Initially, we use 16 particles per cell for electrons and ions in the channel, and 32 particles per cell for electrons and ions in the thin foil.
We limit the simulations to the QED processes that prevail in our conditions, namely gamma-ray generation via nonlinear Inverse Compton scattering and pair production via Bethe-Heitler.

\begin{figure}
    \centering
    \includegraphics[width=0.48\textwidth]{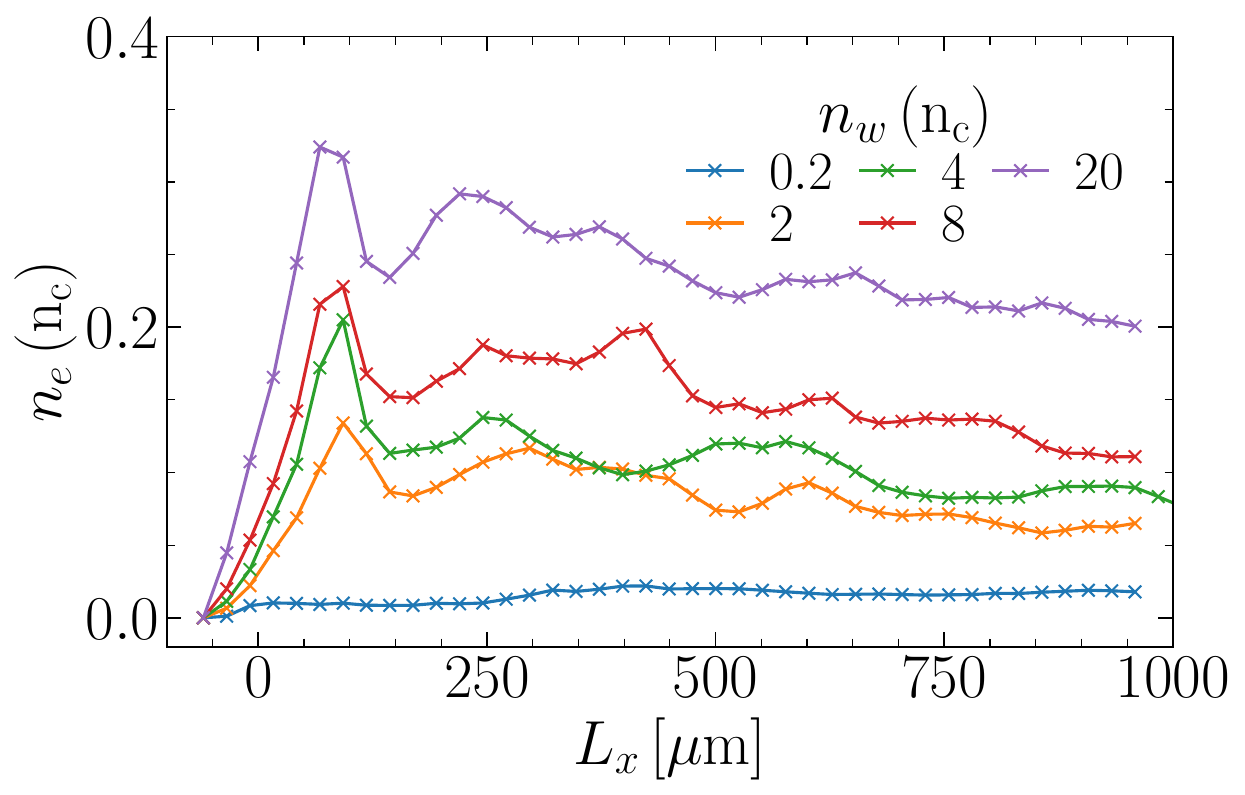}
    \caption{Average density of electrons loaded $n_e$ in the plasma channel as a function of the distance traveled by the peak of the laser in the channel. This distance corresponds to the distance the peak of the laser travels from the entrance to the plasma channel. At each position, it is computed by averaging the electron density in the channel, as illustrated in Figs.~\ref{fig:beam_loading}a)-e). Each curve represents a different initial condition, where the wall density is varied in the range $n_w\in (0.2, \, 2, \, 4, \, 8, \, 20) \times 10^{21} \, \rm cm^{-3}$ and $n_c\simeq 10^{21} \, \rm cm^{-3}$ is the critical density for the laser.
    }
    \label{fig:beam_loading_time_evo}
\end{figure}

\subsection{Laser-driven electron beam loading}

In this section, we illustrate how electron beam loading takes place in PIC simulations, and how it induces fields guiding positrons.
We present the results of five simulations, where all parameters are kept constant and defined above (see also the first row of Table~\ref{tab:simulation_parameters}).
We only vary the wall density of the plasma channel in the range $n_w\in (0.2, \, 2, \, 4, \, 8, \, 20) \times 10^{21} \, \rm cm^{-3}$.

\begin{figure*}
    \centering
    \includegraphics[width=0.96\textwidth]{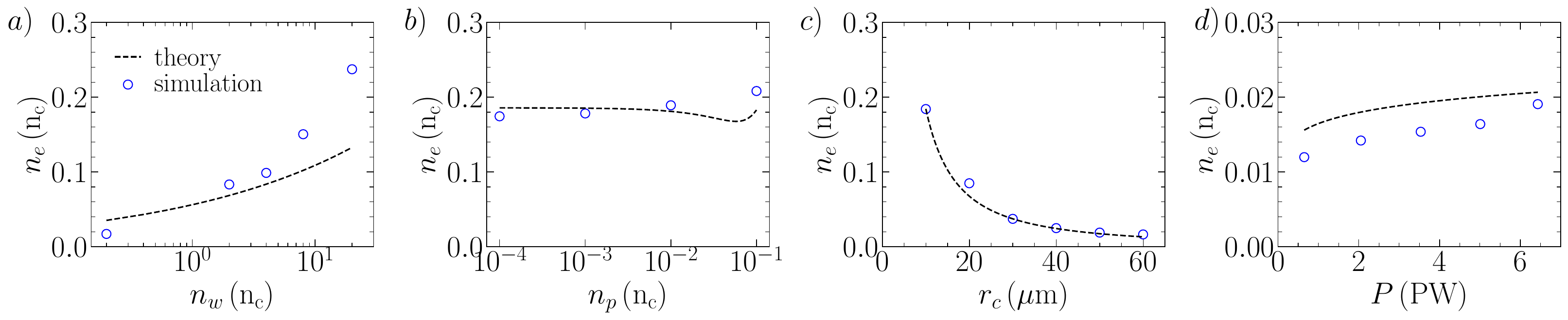}
    \caption{Average density of electrons loaded in the plasma channel from PIC simulations (circles) and our estimate from Eq.~\eqref{eq:density_loaded} (dashed line). Each circle represents the steady-state density of electrons loaded in the channel inferred from a time evolution as illustrated in Fig.~\ref{fig:beam_loading_time_evo}. We vary the channel's wall density in panel a), the channel's central density in panel b), the channel's radius in panel c), and the laser power in panel d). For each panel, the plasma and laser conditions are given in Table~\ref{tab:simulation_parameters}.
    }
    \label{fig:beam_loading_scan}
\end{figure*}

Figure~\ref{fig:beam_loading}a) shows the laser electric field (in red/blue) and the electron density (in green).
At the front of the laser, electrons are expelled by the ponderomotive force and form an ion cavity.
We also observe that electrons are loaded at the rear of the laser.
This is possible by the interplay of the $-e\mathbf{E}$ and $-e\mathbf{v}\times \mathbf{B}$ components.
At this instant, the measured electron charge density is already larger than the background ion charge density.
As a result, the transverse force associated with the static fields generated enables the guiding of positrons.
Indeed, Fig.~\ref{fig:beam_loading}b) shows that in the region where electron beam loading is large, the resulting transverse fields are negative, and therefore the transverse force on positrons is negative, pointing toward the channel axis.
When moving from Fig.~\ref{fig:beam_loading}a) down to Fig.~\ref{fig:beam_loading}e), the wall density in the channel increases in the range $n_w\in (0.2, \, 2, \, 4, \, 8, \, 20) \times 10^{21} \, \rm cm^{-3}$.
The main trend we note is that the amount of electrons loaded increases with the wall density.
This evolution is in line with the injection process from the walls we described in section~\ref{sec:theory}.
The evolution of the guiding fields is seen by moving from Fig.~\ref{fig:beam_loading}f) down to Fig.~\ref{fig:beam_loading}j).
We observe the amplitude of the resulting transverse field on positrons increases with the wall density, due to the increased beam loading.
However, this stronger guiding is achieved to the detriment of a smaller guiding structure.
Indeed, ions are loaded and accelerated at the center of the channel for the highest wall densities considered:$n_w=8 \times 10^{21} \, \rm cm^{-3}$ in Fig.~\ref{fig:beam_loading}h) and $n_w=2 \times 10^{22} \, \rm cm^{-3}$ in Fig.~\ref{fig:beam_loading}j).
The reason is that in these cases, electron beam loading, and therefore the charge separation field mediating ion acceleration, are the largest.

After this illustration of the beam loading process, we are eager to understand quantitatively how electron beam loading scales in realistic PIC simulations.
To achieve this purpose, our goal is to compare the amount of electron beam loading as seen in Fig.~\ref{fig:beam_loading}a) with the theoretical estimate from Eq.~\eqref{eq:density_loaded}.
The latter provides a density averaged on the channel's volume. 
For this reason, we have computed the spatial average electron beam loading density $n_e$ in the plasma channel.
To do this, we consider the instantaneous electron density as in Fig.~\ref{fig:beam_loading}a), subtract the initial electron density profile and average spatially the density below the initial channel's radius of $20 \, \rm \mu m$.
The subtraction method is used to compute the excess charge injected from the channel walls, as they all belong to the same numerical species.
In Fig.~\ref{fig:beam_loading_time_evo}, we plot the evolution of the spatial average electron beam loading density as a function of the laser propagation distance in the channel denoted $L_x$.
Each curve represents one of the simulations from Fig.~\ref{fig:beam_loading}, where the wall density varies in the range $n_w\in (0.2, \, 2, \, 4, \, 8, \, 20) \times 10^{21} \, \rm cm^{-3}$.
For all cases, we observe that the density of electrons loaded increases as the laser propagates further in the channel and seems to saturate after $\sim 250 \, \rm \mu m$ of laser propagation, with relatively small fluctuations.
The presence of a quasi-plateau is in line with our theoretical model, which assumes a constant steady-state value for the average loaded electron beam density.
From the data in Fig.~\ref{fig:beam_loading_time_evo}, we measure the time-average loaded electron beam density (over the interval $L_x>250 \, \rm \mu m$) and report it in Fig.~\ref{fig:beam_loading_scan}a) as a function of the wall density of the PIC simulations performed.
We remark that the density of electrons loaded can increase by one order of magnitude, from $2\times 10^{19} \, \rm cm^{-3}$ to $2\times 10^{20} \, \rm cm^{-3}$ when the wall density is increased by two orders of magnitudes from $2\times 10^{20}  \, \rm cm^{-3}$ to $2\times 10^{22}  \, \rm cm^{-3}$.
This evolution confirms the trend already reported in Fig.~\ref{fig:beam_loading_time_evo} that for a higher wall density, more electrons are loaded in the channel.
We underline that it is quantitatively predicted by our theoretical estimate from Eq.~\ref{eq:density_loaded}.
While this agreement constitutes a first argument in favor of our theoretical model describing electron beam loading, it remains to be complemented by additional simulations.

\begin{table*}
\caption{\label{tab:simulation_parameters}Summary of the initial conditions we chose in the PIC simulations presented in Fig.~\ref{fig:beam_loading_scan}. The columns represent the main parameters of the interaction: the channel's wall density, the channel's central density, the channel's radius and the laser power. Each row represents a scan over one of these parameters, while the others are kept constant.}
\begin{ruledtabular}
\begin{tabular}{ccccc}
&Channel's wall density&Channel's central density&Channel's radius&Laser power\footnote{In all simulations, the electric field  has a Gaussian temporal profile of $150 \, \rm fs$ (FWHM).}\\
&$n_w \, (10^{21} \, \rm /cm^3)$& $n_p \, (10^{21} \, \rm /cm^3)$&$r_c \, \rm (\mu m) $&$P \, \rm (PW)$\\ \hline
Fig.~\ref{fig:beam_loading_scan}a)&$0.2,\, 2,\, 4,\, 8,\, 20$&$10^{-3}$&$20$&$9$\footnote{For all these scans, the electric field has a Gaussian transverse profile of $5.3 \, \rm \mu m$ (FWHM).} \\
Fig.~\ref{fig:beam_loading_scan}b)&$2$&$10^{-4}, \, 10^{-3}, \,10^{-2}, \,10^{-1}$&$10$&$9$\footnotemark[2]\\
Fig.~\ref{fig:beam_loading_scan}c)&$2$&$3\times 10^{-3}$&$10, \, 20, \, 30, \, 40, \, 50, \, 60$&$9$\footnotemark[2]\\
Fig.~\ref{fig:beam_loading_scan}d)&$2$&$10^{-3}$&$40$&$0.6, \, 2, \, 3.5, \, 5, \, 6.5$\footnote{For this parametric scan, the envelope of the laser electric field  has a Gaussian transverse profile of $10.6 \, \rm \mu m$ (FWHM).}\\
\end{tabular}
\end{ruledtabular}
\end{table*}

We ran complementary PIC simulations in order to further test the validity of our theoretical estimate for electron beam loading.
For the sake of clarity, we summarise the main parameters for each of the 20 PIC simulations we conducted in Table~\ref{tab:simulation_parameters}.
We vary four characteristic quantities, one at a time.
We consider different wall and central densities for the plasma channel, as well as various channel's radii and laser powers in the Peta Watt range.
In Figs.~\ref{fig:beam_loading_scan}a)-d), the overall comparison between the density of electrons loaded measured in PIC simulations and the theoretical estimate is correct.
For the four parameters scanned, the qualitative evolution of beam loading is well-reproduced by Eq.~\ref{eq:density_loaded}.
We therefore conclude that our estimates of electron beam loading provides the correct scaling laws as a function of the main plasma and laser parameters.

The agreement between our theoretical scaling of electron beam loading with PIC simulations is obtained despite complex interaction conditions, the wide range of laser-plasma parameters and nonlinear plasma processes involved.
It relies on one main hypothesis, that the electrons are extracted from the walls of the plasma channel, at a radius approximated to be the blowout radius.
In the next section, we test the validity of this assumption directly from the data of our PIC simulations.

\subsection{Injection of electrons from the walls of the channel}

In this section, we analyse the phase space of electrons loaded in the plasma channel.
Our goal is to investigate whether they are effectively injected from the walls in PIC simulations, as we assumed in the theoretical modelling.

\begin{figure}
    \centering
    \includegraphics[width=0.48\textwidth]{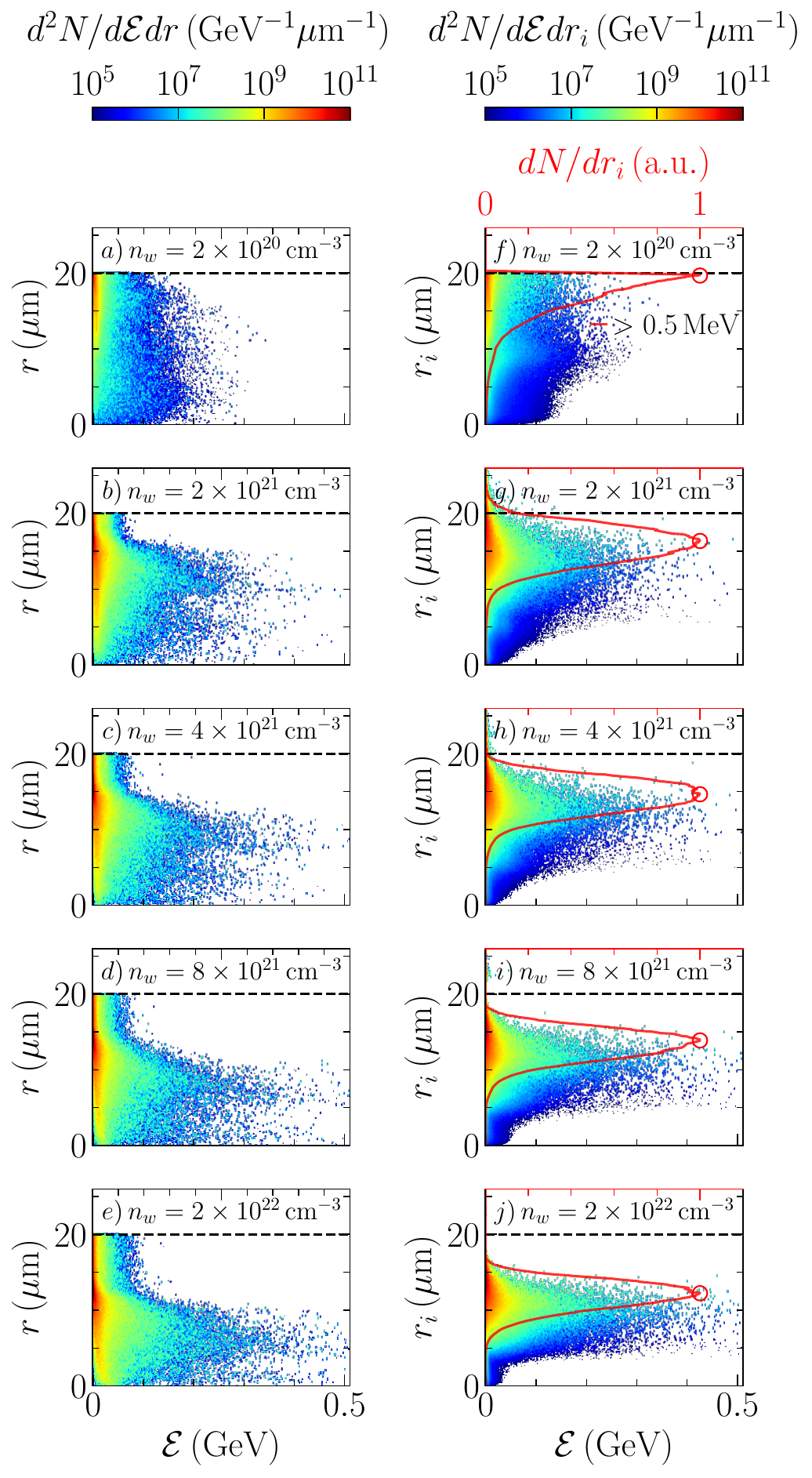}
    \caption{Electron injection as a function of the wall density of the channel. Panels a)-e) illustrate the $(r,\mathcal{E})$ phase space of electrons. Panels f)-j) depict the $(r_i,\mathcal{E})$ phase space of electrons, where $r_i$ is the \emph{initial} radius of electrons when they are injected. The red curve shows the distribution in initial radius of electrons.  When moving from the top row to the bottom row, we increase the wall density of the plasma channel in the range $n_w\in (0,2, \, 2, \, 4, \, 8, \, 20) \times 10^{21} \, \rm cm^{-3}$. The data is shown after the peak of the laser has propagated over a distance $L_x=200 \, \rm \mu m$ in the channel.
    }
    \label{fig:radius_injection}
\end{figure}

The phase space of electrons is displayed in Fig.~\ref{fig:radius_injection}, for simulations where we modify the wall density of the plasma channel.
The exact simulation parameters are in the first row of Table~\ref{tab:simulation_parameters}, and are the same as in Fig.~\ref{fig:beam_loading}.
We first focus on one simulation, where the wall density is the smallest: $n_w=2\times 10^{20} \, \rm cm^{-3}$, see Figs.~\ref{fig:radius_injection}a)f).
Figure~\ref{fig:radius_injection}a) shows the 2D distribution $d^2N/d\mathcal{E}dr$ for all electrons in the channel.
We recall that $r$ is the radial coordinate of electrons ($r^2=y^2+z^2$) and $\mathcal{E}$ is their kinetic energy.
We observe that relativistic electrons reach energies in the range of hundreds of MeV at the instant represented, after $200 \, \rm \mu m$ of laser propagation in the channel.
This figure only gives information on the instantaneous radius of electrons $r$, but not on their initial radius, denoted $r_i$, from where they are injected in the channel.
From the simulation data, we have inferred the initial distance from the axis $r_i$ of each electron and computed numerically the 2D distribution $d^2N/d\mathcal{E}dr_i$, that is plotted in Fig.~\ref{fig:radius_injection}f).
We note that the initial radius of electrons seems to coincide with the channel's radius $r_i\simeq r_c = 20 \, \rm \mu m$.
To ensure this quantitatively, we integrate the 2D distribution $d^2N/d\mathcal{E}dr_i$ to obtain a distribution in initial radius labeled $dN/dr_i$, see the red curve in Fig.~\ref{fig:radius_injection}f).
The peak of the distribution $dN/dr_i$ is marked by a red circle and is located at a radius $r_i\simeq r_c = 20 \, \rm \mu m$.
This means that for these parameters and at this instant, most of the loaded electrons are injected from the channel wall itself.
Note that we separate loaded from unloaded electrons with the energy criteria ($>mc^2$), as they all belong to the same numerical species.
As we observe from Fig.~\ref{fig:radius_injection}a) down to Fig.~\ref{fig:radius_injection}e), the instantaneous phase space of electrons indicates that they reach hundreds of MeV for all wall densities at stake.
In parallel, when moving down from Fig.~\ref{fig:radius_injection}f) to Fig.~\ref{fig:radius_injection}j), we can observe how the injection radius of electrons varies when increasing the wall density in the plasma channel in the range $n_w\in (0.2, \, 2, \, 4, \, 8, \, 20) \times 10^{21} \, \rm cm^{-3}$.
This radius of injection decreases from $r_i\simeq 20 \, \rm \mu m$ down to $ \simeq 14 \, \rm \mu m$ when increasing the wall density from $2 \times 10^{20} \, \rm cm^{-3}$ to $2 \times 10^{22} \, \rm cm^{-3}$, as pinpointed by the red circles.

\begin{figure}
    \centering
    \includegraphics[width=0.48\textwidth]{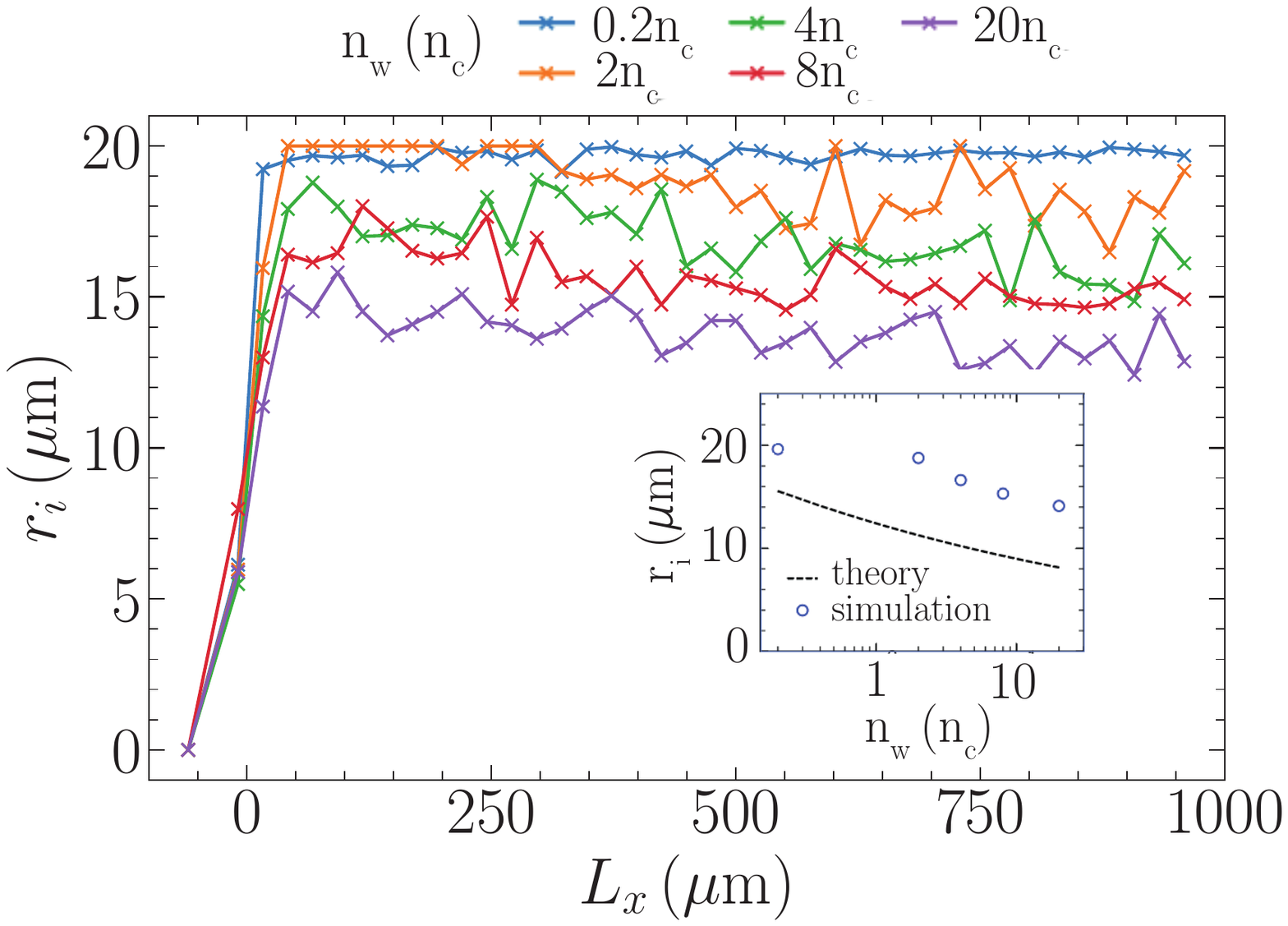}
    \caption{Injection radius $r_i$ of electrons loaded in the plasma channel as a function of the distance travelled by the peak of the laser in the channel. At each position, it is computed from the peak of the distribution in initial radius $dN/dr_i$ from Fig.~\ref{fig:radius_injection}f)-j). Each curve represents a different initial condition, where the wall density is varied in the range $n_w\in (0.2, \, 2, \, 4, \, 8, \, 20) \times 10^{21} \, \rm cm^{-3}$ and $n_c\simeq 10^{21} \, \rm cm^{-3}$ is the critical density for the laser.
    }
    \label{fig:radius_injection_time_evo}
\end{figure}

We will now investigate how these values compare with what we expect from the analytical model. The injection radius of electrons loaded seems to fluctuate as the laser propagates in the channel.
In Fig.~\ref{fig:radius_injection_time_evo}, we plot it for the different simulations presented in Fig.~\ref{fig:radius_injection}, where we modify the wall density in the channel.
We stress that the value reported is the peak of the distribution in of the initial radial position of injected electrons, as we computed it in Figs.~\ref{fig:radius_injection}f)-j) (see the red distribution function and the circle).
It represents the initial radius of \emph{most} electrons, but not all of them.
After the acceleration stabilizes, the average $r_i$ of the injected electrons is constant, which is expected as the fresh plasma always has the same plasma profile so the injection is the same beyond the initial plasma ramp.
For each simulation, we determine this quasi-constant value by averaging the evolution from Fig.~\ref{fig:radius_injection_time_evo} over the interval $L_x>250 \, \rm \mu m$.
The result of this analysis is displayed in the inset of Fig.~\ref{fig:radius_injection_time_evo} along with the theoretical estimate for the injection radius of electrons determined from Eq.~\eqref{eq:radiusinjection}.
Although we note the theory predicts a slightly lower numerical value than the simulation data, the qualitative trend it provides is correct.
This trend from the theory points that for a larger wall density, the amplitude of the perpendicular electrostatic field built to balance the ponderomotive push is larger, and the blowout radius thus decreases.
As a result, the laser is more confined to the center of the channel, and it injects electrons at a smaller radius than $r_i$ in the analytical model.
The theory neglects the dynamical evolution of the laser and the ion motion, so a perfect agreement is not expected. Nevertheless, it seems that this model gets quite close to fully describing the injection process. 

In this section, we have confirmed the main hypothesis of our theoretical model on electron beam loading using PIC simulations.
Indeed we demonstrated that electrons are injected from the walls of the plasma channel for various laser and plasma conditions.
The main outcome is that our model correctly describes the injection process of electrons in the channel, and how it scales with the plasma channel and laser profiles.

\subsection{Evidences of Direct Laser Acceleration}

We now focus on the dynamics of positrons and prove that they experience Direct Laser Acceleration in PIC simulations.
We include the analysis of individual trajectories as well as the whole positron spectrum.

\begin{figure}
    \centering
    \includegraphics[width=0.48\textwidth]{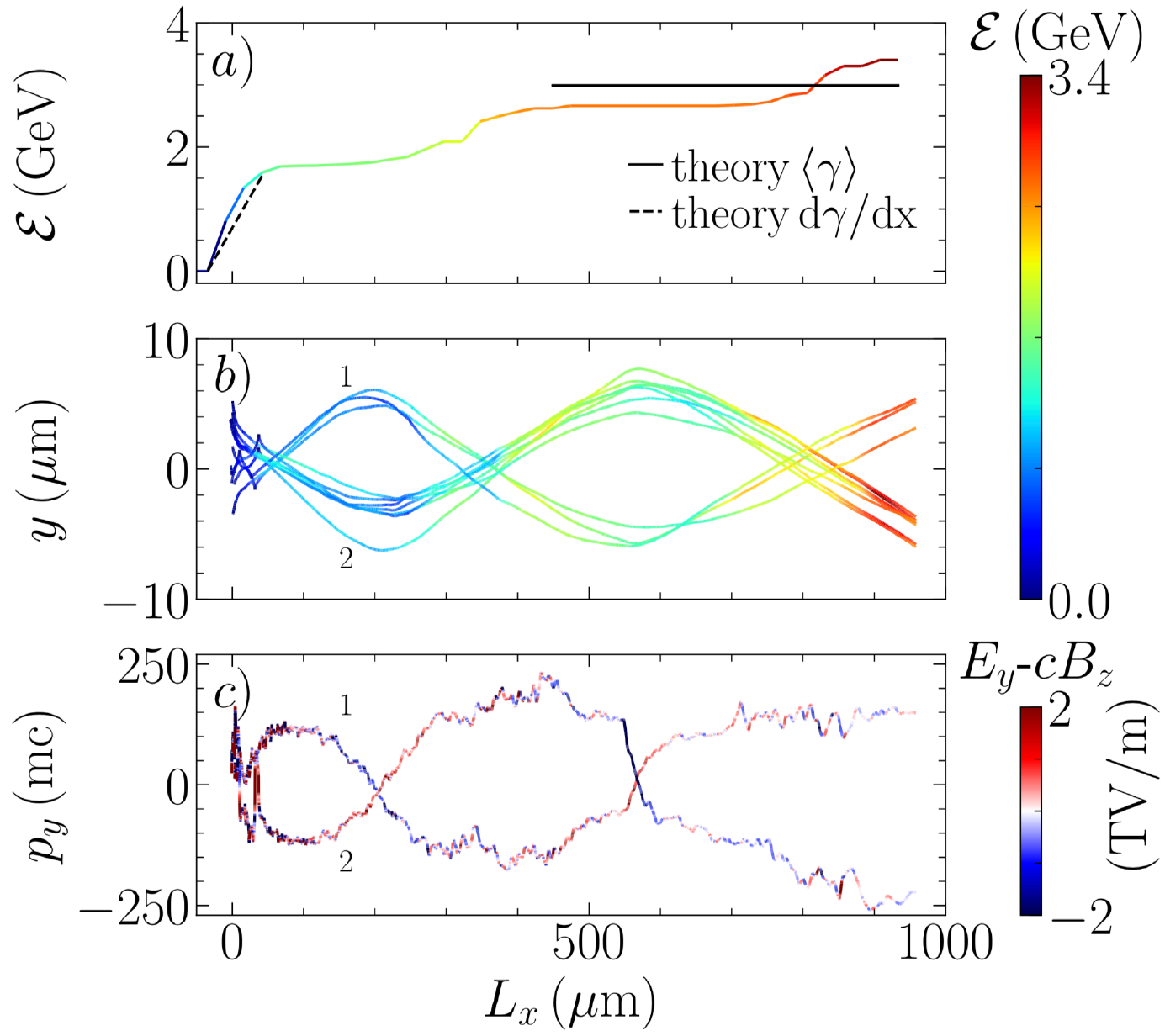}
    \caption{a) Maximum positron energy in the simulation domain as a function of laser propagation distance in the channel with the theoretical estimates from Eqs.~\eqref{eq:scaling_gamma}-\eqref{eq:scaling_gradient}. b) Oscillations of two types of representative positron trajectories in the ($x-y$) plane, where $y$ is the laser polarisation direction and $x$ the laser propagation direction. c) Evolution of the transverse positron momentum $p_y$ and the transverse fields experienced by the two types of positron trajectories. The transverse fields are the ones of the plasma channel (mode $m=0$) and do not include the laser's contribution (mode $m=1$).
    }
    \label{fig:tracks_gamma_field} 
\end{figure}

Positron acceleration is depicted in Fig.~\ref{fig:tracks_gamma_field}.
We consider the simulation with parameters described in Sec.~\ref{sec:sim_params}, where the central channel density is $n_p=10^{18} \, \rm cm^{-3}$, the wall density is $n_w=4\times10^{21} \, \rm cm^{-3}$, the channel's radius is $r_c=20 \, \rm \mu m$ and the laser power $9 \, \rm PW$.
The maximum positron energy in the simulation domain is plotted in Fig.~\ref{fig:tracks_gamma_field}a), as a function of the laser propagation distance in the channel.
We see the instantaneous energy gain rate is as high as $\simeq 15 \, \rm TeV/m$ in the first $100 \, \rm \mu m$ of interaction.
This transient value is unusually large, even for plasma-based accelerators.
This can be explained by the fact that the laser field amplitude reaches $\simeq 10^{23} \, \rm Wcm^{-2}$ due to strong self-focusing at the entrance of the channel.
The maximum energy achieved for positrons is $3.4 \, \rm GeV$ after 1 millimeter of propagation in the channel, corresponding to an average energy gain rate of $3.4 \, \rm TeV/m$.
We note that both the peak energy and the instantaneous energy gain rate are well-estimated by our analytical scaling laws from Eq.~\eqref{eq:scaling_gamma} and Eq.~\eqref{eq:scaling_gradient}, where we account for radiation reaction with Eq.~\eqref{eq:I_rr}.
However, this agreement is challenging to obtain solely from theoretical assumptions as the scaling law has a non-linear dependence with respect to three parameters: the laser intensity, the electron beam loading and the initial transverse position of the accelerated positrons.
Since it is not yet possible to infer these quantities from theory, we have used the values measured in the simulation.
For instance, the agreement on the energy gain rate is obtained on the first $100 \, \rm \mu m$ of propagation for a measured laser intensity of $10^{23} \, \rm Wcm^{-2}$ and a measured beam loading of $n_e=4\times10^{19} \, \rm cm^{-3}$.
The maximum energy is reached after $500  \, \rm \mu m$ of propagation, when the measured laser intensity has reduced to $2\times 10^{22} \, \rm Wcm^{-2}$ and the beam loading has grown larger to $n_e=10^{20} \, \rm cm^{-3}$.
This steady-state value of electron beam loading of $n_e=10^{20} \, \rm cm^{-3}$ is very close to the theoretical estimate of Eq.~\eqref{eq:density_loaded} that provides $n_e=9\times 10^{19} \, \rm cm^{-3}$, as seen in Fig.~\ref{fig:beam_loading_scan}a).
Finally, both agreements with theory are obtained considering that positrons are injected with an initial radius of $5.3 \, \rm \mu m$, namely the FWHM of the electric field and that the phase velocity is equal to the speed of light.

In Quasi-3D PIC simulations, resonant positrons oscillate as expected from the theoretical description.
Figure~\ref{fig:tracks_gamma_field}b) details how the transverse coordinate $y$ of high-energy positrons evolves as a function of their propagation in the channel, where $y$ is the laser polarisation direction.
First they experience fast oscillations in the laser field for a propagation distance $L_x<100 \, \rm \mu m$.
Beyond this limit, the frequency of the oscillations decreases and we measure a period of $\simeq 0.8 \, \rm mm$.
Looking at the first type of trajectory indexed by 1 in Fig.~\ref{fig:tracks_gamma_field}b), we observe positrons are guided along the axis and confined to the center of the channel ($ y\leq 10 \, \rm \mu m$) for as long as 1 millimeter.
At $L_x=200 \, \rm \mu m$ of propagation, the positrons turn toward the center of the channel instead of being expelled.
This behavior is explained in Fig.~\ref{fig:tracks_gamma_field}c), where we represent the transverse momentum of a representative positron $p_y$ and the quasi-static (mode $m=0$) transverse channel fields it experiences $E_y-cB_z$.
We note the transverse momentum decreases rapidly from $p_y/mc\simeq 120$ to $p_y/mc\simeq -120$ in a distance of $150 \, \rm \mu m$.
This momentum decrease is quantitatively compatible with the negative self-generated fields from the plasma channel with an amplitude that reaches $E_y-cB_z\simeq 1 \, \rm TV/m$.
We can distinguish these quasi-static fields from the laser field thanks to the azimuthal mode decomposition of our simulations.
At a later propagation distance ($L_x\simeq 600 \, \rm \mu m$), we see in Fig.~\ref{fig:tracks_gamma_field}b) the same positrons are focused again toward the laser propagation axis.
This can be explained as their momenta increase due to the now positive transverse channel fields.
For the sake of the example, we also added another type of trajectory labeled "2" that oscillates resonantly.
The transverse momentum of this other test-positron changes sign, and this can be as well explained by the strong transverse plasma fields.

\begin{figure}
    \centering
    \includegraphics[width=0.48\textwidth]{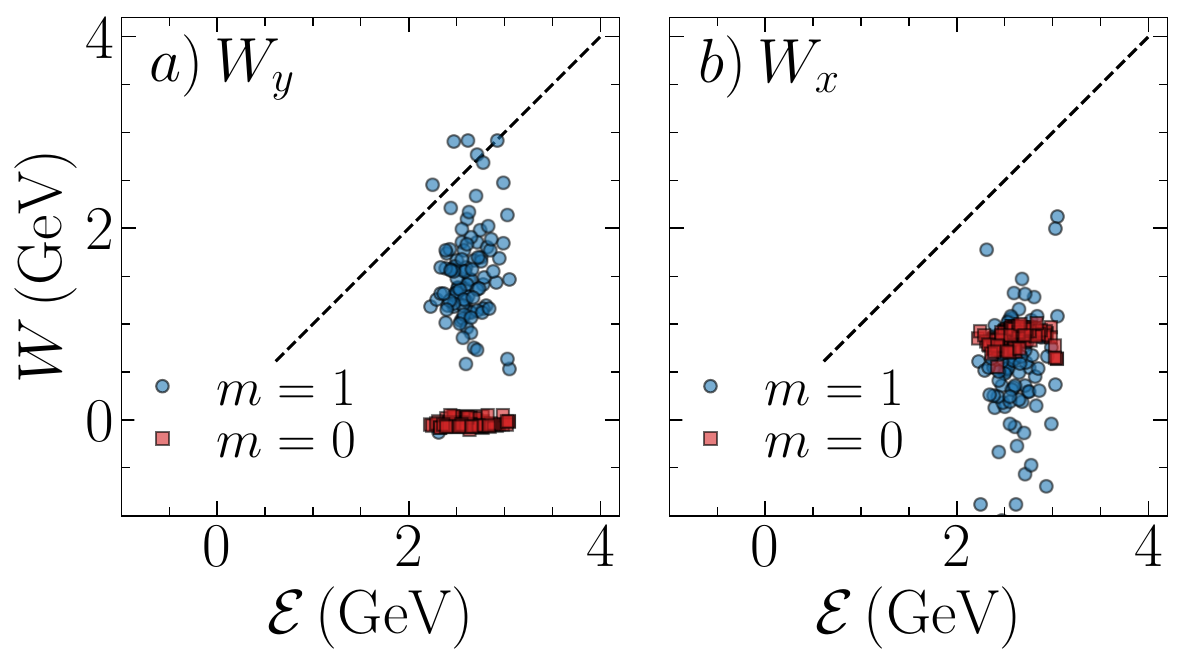}
    \caption{Energy gain of positrons experiencing Direct Laser Acceleration accumulated on 1 mm of laser propagation in the channel. The energy gains $W$ represent the accumulated work in the channel fields ($m=0$, red squares) and the laser fields ($m=1$, blue circles) after a propagation distance of $1 \, \rm mm$ in the plasma channel. We recall that the laser propagation direction is $x$, and the polarisation direction is $y$.
    }
    \label{fig:work_on_posi}
\end{figure}

After this analysis of representative trajectories, we consider the hundred most energetic positrons and prove they experience Direct Laser Acceleration.
To this purpose, we illustrate the accumulated work performed by different field components on positrons, over the whole propagation of the laser pulse in the channel ($1 \, \rm mm$).
The energy gain $W$ is decomposed in two directions $W_{y/x}=\int \mathbf{v}\cdot \mathbf{E}_{y/x} \, dt$, where $y$ is the laser polarisation direction and $x$ the laser propagation direction.
We also separate the contribution of the channel fields (mode $m=0$) and the laser (mode $m=1$).
A first look at Fig.~\ref{fig:work_on_posi} shows the gain from the laser field in its polarisation direction ($W_y$ for mode $m=1$ as blue circles) prevails over the gain in plasma wakefields ($W_x$ for mode $m=0$ as red squares).
Despite their smaller contribution, the role of plasma wakefields is non-negligible.
In addition, the contribution of the longitudinal laser field component ($W_x$ for mode $m=1$ as blue circles) is also of comparable amplitude.
We conclude that a complete modeling of Direct Laser Acceleration should include these contributions.
This effort is already discussed~\cite{PoPNaseri2010,PoPWang2019} at lower laser intensities ($\sim 10^{19} \, \rm Wcm^{-2}$), but is not yet generalized for the intensities we consider ($\sim 10^{22} \, \rm Wcm^{-2}$).

\begin{figure}
    \centering
    \includegraphics[width=0.48\textwidth]{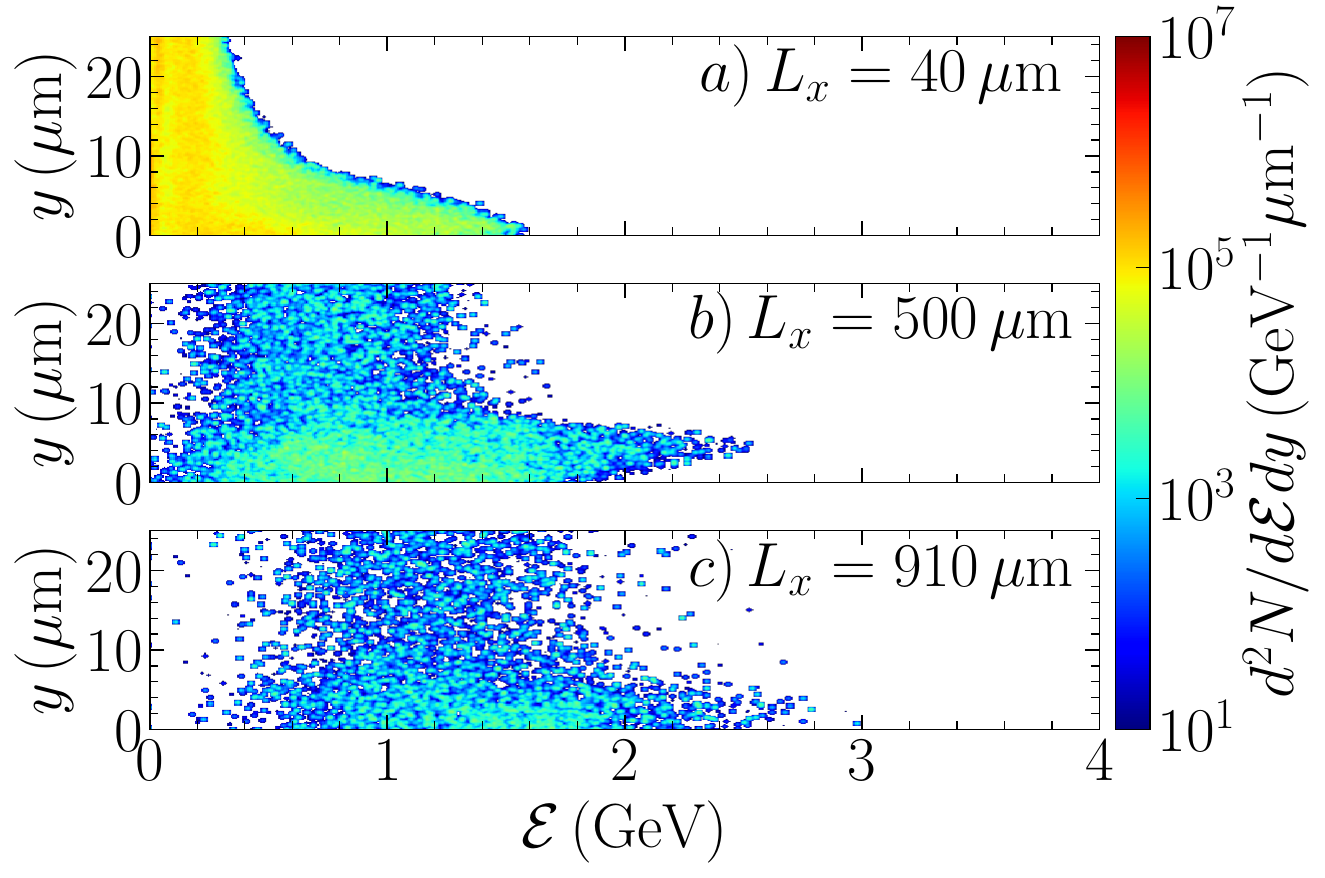}
    \caption{Phase space $(y-\mathcal{E})$ of positrons. It is represented just after their creation in the thin foil in panel a), after $500 \, \rm \mu m$ of laser propagation in panel b) and after $910 \, \rm \mu m$ of laser propagation in panel c).
    }
    \label{fig:spectrum_posi}
\end{figure}

The positron spectrum simulated contains an unambiguous proof they experience DLA.
We show the $(y,\mathcal{E})$ phase space of positrons in Fig.~\ref{fig:spectrum_posi} at three instants, where $y$ is the laser polarisation direction.
At the beginning of the interaction $L_x=40 \, \rm \mu m$ in Fig.~\ref{fig:spectrum_posi}a), the laser has just travelled through the thin foil and enters the channel.
At this point, positrons reach a maximum energy of $1.5 \, \rm GeV$ and it is achieved by the positrons located on the laser propagation axis $y=0$.
This gain is solely due to positrons oscillating in the laser field as it goes through the thin foil in the Relativistically Self-Induced Transparency regime.
At a later propagation distance of $L_x=500 \, \rm \mu m$ in Fig.~\ref{fig:spectrum_posi}b), the maximum energy is now $2.5 \, \rm GeV$.
It is reached by positrons that are not on the laser propagation axis ($y=0$), but slightly off-axis for a transverse position $y\simeq 5 \, \rm \mu m$.
This behavior comes from the amplitude of the betatron oscillations of the positrons of $\simeq 5-10 \, \rm \mu m$, as seen in Fig.~\ref{fig:tracks_gamma_field}b).
This so-called forking structure has already been observed experimentally for electrons and is used as an evidence of the DLA mechanism~\cite{PPCFShaw2018}.
Our simulations confirm that the same type of evidence can be expected for positrons.
We also highlight that the central beam energy is  $\sim 1 \, \rm GeV$.
The final inset in Fig.~\ref{fig:spectrum_posi}c) displays the positron spectrum close to the end of the plasma channel for  $L_x=910 \, \rm \mu m$.
The maximum energies are higher $\simeq 3 \, \rm GeV$, while the central beam energy kept increasing to $1.5 \, \rm GeV$.
However, this increase is done to the detriment of beam charge.
Indeed, positrons are continuously lost over the propagation of the laser in the channel due to an imperfect guiding.
We have quantified the loss of positrons for these laser and plasma parameters. The  Tab.~\ref{tab:conversion_eff} provides the initial laser energy for ELI-Beamlines at 9 PW, the total energy of all the positrons created in the foil, and the energy of the positrons and electrons in the beam after 1 mm of propagation.
The $1.3 \, \rm kJ$ laser produces $2\times 10^6$ positrons in total in the thin foil, with a total energy of $42 \, \rm \mu J$ at the time of creation.
After 1 mm of propagation in the plasma channel, there are $10^4$ positrons left in the beam, for a total energy of $3 \, \rm \mu J$ (these represent about $0,5~\%$ of initial positrons, so their initial energy content was $\sim 0.2~\rm \mu J$, while the rest was obtained through acceleration).
One can define an energy conversion efficiency from the laser to the positrons created in the foil $10^{-6} \%$, which may be improved by optimizing the laser and foil parameters in the future.
The total energy contained in the positron  beam after 1 mm represents $\simeq 7\%$ of the energy converted to BH positrons within the target.

\begin{table}
\caption{\label{tab:conversion_eff}Energy of the incident laser (column 1), of all the positrons created in the foil (column 2), of the positrons in the beam (column 3) and of the electrons in the beam (column 4). The energies of electrons and positrons in the beam are given after 1 mm of laser propagation in the plasma channel.}
\begin{ruledtabular}
\begin{tabular}{ccccc}
&Laser&$e^+$ (foil)&$e^+$ (beam)&$e^-$ (beam)\\ \hline
Energy&$1.3 \, \rm kJ$&$42 \, \rm \mu J$&$3 \, \rm \mu J$&$4 \, \rm J$ \\
Efficiency ($\%$)&$100$&$3.2\times 10^{-6}$&$3.0\times 10^{-7}$&$0.3$ \\
\end{tabular}
\end{ruledtabular}
\end{table}

To conclude this section, we brought evidence that Direct Laser Acceleration of positrons can take place using Quasi-3D PIC simulations.
This confirmation is done through the analysis of single trajectories and also the full positron spectrum.

\section{Discussion}

We will now compare our approach with existing studies documented in the literature.
While post-acceleration configurations offer better positron charge and beam quality, they lack compactness as they require a kilometer-long accelerator~\cite{NatCorde2015,NCGessner2016,SRDoche2017,PRLLindstrom2018,PRLJain2015,PRLVieira2014,PPCFLi2019,PRRHue2021,PRLWang2008,PRSTABKimura2011,SRYi2014,PRABDiederichs2019,PRLSilva2021,PREReichwein2022}.
The distinct advantage of our method lies in its ability to generate and accelerate positrons to a GeV level with a single 10 PW laser pulse.
Other schemes utilizing multi-PW lasers either solely concentrate on pair creation without acceleration~\cite{PRLSokolov2010,PRABLobet2017,PRABlackburn2017} or employ a 2D geometry that hinders accurate quantitative predictions~\cite{PRLChen2010, PPCFYan2017,SRVranic2018,CPHe2021}.
Despite this observation, it is worth underlining a recent experiment related to the creation of a high-quality positron beam~\cite{SRStreeter2024}, marking a milestone toward laser-driven positron acceleration in a plasma wakefield.
Our Quasi-3D approach enables us to obtain precise quantitative estimates for our positron generation and acceleration scheme.
We remark that DLA of positrons was observed in one numerical work~\cite{CPHe2021}.
However, we bring an original contribution by introducing a theoretical framework describing this process, along with  validation against PIC simulations.  

We have also assessed the experimental feasibility of our setup.
We considered nominal parameters for the ELI laser pulse~\cite{RLECheriaux2018}, with a total energy of $1.3 \, \rm kJ$ and a peak power of $9 \, \rm PW$.
There is no particular constraint of the thin Aluminum foil, except that it should be thin enough ($\simeq 240 \, \rm nm$) to enable a significant laser transmission.
The preformed plasma can be produced with commercially available plasma jet~\cite{RSISylla2012}, with a density of $10^{21} \, \rm cm^{-3}$ and a  length that can be as short as $400 \, \rm \mu m$.
For instance, a dense and narrow plasma channel was formed in an experiment related to ion acceleration in near-critical-density plasmas with 0.1-1 PW-class lasers~\cite{EPSOspina2023}.
Although the channel is not fully characterized, its wall density is on the order of $10^{20} \, \rm cm^{-3}$, its length is $800 \, \rm \mu m$ and its radius is $150 \, \rm \mu m$.
These numbers are comparable to what we are requiring for our setup, paving the way for a future experiment.
In the context of ELI-Beamlines, the channel could be formed with the 1 PW laser system (HALPS), while the positron acceleration could be done with the 10 PW system (ATON).

\section*{Conclusions}

To conclude, we have investigated Direct Laser Acceleration of Bethe-Heitler positrons in laser-channel interactions.
In this setup, positrons are created in a thin aluminum foil and accelerated in a preformed plasma channel by the same high-power laser pulse (10 PW).

We first build a theoretical framework of electron beam loading during the propagation of a high-power laser in a dense plasma channel.
Our main assumption is that electrons are injected from the walls of the plasma channel to its center via the interplay of the two laser field components.
In light of this, we derived the density of electrons loaded as a function of the diverse parameters of the plasma channel and the laser power.
In addition, we use this result to introduce a theoretical framework for DLA of positrons.
The main outcome is that we obtain a quantitative estimate of electron beam loading as a function of the initial laser and plasma conditions, thus gaining a better understanding of the exact acceleration conditions of positrons.

We later confirm the assumptions and predictions of this theoretical model with Quasi-3D PIC simulations.
In particular, our estimate for electron beam loading predicts the correct scaling measured from Quasi-3D PIC simulations where we vary a large range of laser and plasma conditions.
We have also checked that electrons are actually injected from the walls of the plasma channel directly in PIC simulations, as assumed in our model.
Finally, we analyzed single trajectories of positrons and their full spectrum and found several pieces of evidence they experience DLA.

This work directly contributes to advance our theoretical understanding of plasma-based positron acceleration.
It also contributes to lay the bases of a potential experiment at ELI-Beamlines to test a new positron acceleration process.
As a potential prospect, the jet produced may be used to mimic the propagation of astrophysical jets from Gamma Ray Bursts in the laboratory frame, as it contains three species found in fireball jets:  electrons, positrons and x-rays.

\begin{acknowledgments}
We acknowledge the support of the Portuguese Science Foundation (FCT) Grant No. CEECIND/01906/2018,  PTDC/FIS-PLA/3800/2021 DOI: 10.54499/PTDC/FIS-PLA/3800/2021 and UI/BD/151560/2021 DOI:10.54499/UI/BD/151560/2021
We acknowledge PRACE for awarding us access to MareNostrum at BSC (Spain), EuroHPC for awarding us access to LUMI-C at CSC (Finland) and FCT for awarding us access to Oblivion at University of Evora (Portugal).
\end{acknowledgments}

\providecommand{\noopsort}[1]{}\providecommand{\singleletter}[1]{#1}%

\end{document}